\documentclass[twocolumn,tighten]{aastex631}

\usepackage{amsmath}
\usepackage{xspace}
\usepackage{multirow}
\usepackage{fancyapj}
\usepackage{mathtools}
\usepackage{xcolor}

\makeatletter
\def\restartappendixnumbering{\global\applettertrue
\setcounter{table}{0}
\setcounter{figure}{0}
\setcounter{equation}{0}
\def\thetable{\thesection\the\c@table}%
\renewcommand{\theHtable}{Supplement.\thetable}
\def\fnum@table{{\bf\tablename~\thetable}}%
\def\thefigure{\thesection\the\c@figure}%
\def\fnum@figure{{\bf\figurename~\thefigure}}%
}%
\makeatother

\renewenvironment{quote}%
  {\list{}{\leftmargin=0.1in\rightmargin=0.1in}\item[]}%
  {\endlist}

\expandafter\def\csname editcolor1\endcsname{magenta}
\expandafter\def\csname editcolor2\endcsname{red}

\newcommand{\E}[1]{\ensuremath{\times 10^{#1}} }

\newcommand{\kev}{\rm\,keV\xspace}
\newcommand{\hz}{\rm\,Hz\xspace}

\newcommand{\km}{\rm\,km\xspace}

\newcommand{\kpc}{\rm\,kpc\xspace}
\newcommand{\s}{\rm\,s\xspace}

\newcommand{\per}[1]{\rm\,#1\ensuremath{^{-1}}\xspace}
\newcommand{\persq}[1]{\rm\,#1\ensuremath{^{-2}}\xspace}
\newcommand{\lumcgs}{\rm\,erg\per{s}\xspace}

\newcommand{\nicer}{\textrm{NICER}\xspace}

\newcommand{\src}{IGR~J17062\xspace}
\newcommand{\srcfull}{IGR~J17062--6143\xspace}

\definecolor{deeppurple}{RGB}{206, 0, 187}

\begin{document}

\title{The stochastic X-ray variability of the accreting millisecond pulsar IGR J17062--6143}

\author[0000-0002-7252-0991]{Peter Bult}
\affiliation{Department of Astronomy, University of Maryland,
  College Park, MD 20742, USA}
\affiliation{Astrophysics Science Division, 
  NASA Goddard Space Flight Center, Greenbelt, MD 20771, USA}

\begin{abstract}
This work presents an investigation of the stochastic X-ray variability from the $164$ Hz accreting millisecond pulsar IGR J17062--6143, based on regular observations collected with the Neutron Star Interior Composition Explorer between 2017 July and 2020 August. Over this period, the power density spectrum showed a stable morphology, with broad $\sim25\%$ rms band-limited noise below $16\hz$. Quasi-periodic oscillations (QPOs) were occasionally observed, with the most notably detections including a low-frequency QPO centered at $2.7\hz$ and a sharp QPO centered at $115\hz$ that may be a 2:3 resonance with the spin frequency.
Further, the energy dependence of the band-limited noise is studied through a spectroscopic analysis of the complex covariance in two frequency intervals. It is found that the power law continuum is the primary driver for the observed variability, although the thermal (blackbody) emission also appears to be intrinsically variable in area ($5\%$ rms) and temperature ($1\%$ rms). Notably, the $1\kev$ emission feature seen in all X-ray spectra of IGR J17062--6143 varies with the same amplitude as the power law, but systematically lags behind that continuum emission. These results appear consistent with a scenario in which a time variable Compton-scattering corona is the primary source for the observed stochastic variability, with the variability observed in the emission feature and at the lowest photon energies being due to the disk reflection of the power law continuum.
\end{abstract}


\section{Introduction}
\label{sec:intro}

Accreting millisecond X-ray pulsars (AMXPs) are rapidly rotating neutron stars
that accrete matter from a low-mass binary companion star \citep[see][for
a recent review]{DiSalvo2020}. These objects are characterized by their
coherent pulsations; a nearly sinusoidal modulation of the X-ray emission that
directly tracks the neutron star rotation. Through the study of these
pulsations one can extract information on neutron star properties, such as
mass, radius, or magnetic field strength \citep{Poutanen2003, Leahy2008,
Psaltis2014}. Additionally, pulsations can be used to study the torque exchange
between the neutron star and the accretion disk \citep{Bildsten1998b,
Psaltis1999b}, and investigate the evolution of the stellar binary
\citep{Nelson2003}. 

While the pulsar properties of AMXPs alone make them interesting targets for
study, these systems generally also exhibit the phenomenology seen in
non-pulsating low-mass X-ray binaries (LMXBs), including accretion rate
dependent state transitions \citep{Straaten2005}, quasi-periodic variability
(QPOs) \citep{Wijnands1998b, Straaten2005}, and thermonuclear X-ray bursts
\citep{Zand1999, Chakrabarty2003}. Through this combination of pulsar and LMXB
properties the different observables of AMXPs offer complementary views of the
same underlying accretion processes \citep{Wijnands2003, Chakrabarty2003,
Bult2015, Bult2017b}.

The X-ray transient \srcfull (simply \src henceforth) is an especially
interesting system for testing our understanding of accretion onto a weakly
magnetized neutron star. The source was first discovered in Jan/Feb of 2007
\citep{Churazov2007} and has since remained visible at a comparatively low
luminosity of $L_X \approx 5\E{35}\lumcgs$ \citep{Degenaar2017,
Eijnden2018, Bult2021b}. During this unusually long outburst, \src has shown
three energetic intermediate duration type I X-ray bursts \citep{Degenaar2013,
AtelNegoro15, AtelNishida20a}, but has otherwise remained remarkably stable. 
Previous studies have investigated its type I X-ray bursts and their interactions with
the accretion environment \citep{Degenaar2013, Keek2017, Bult2021c}, the X-ray
spectroscopy \citep{Degenaar2017, Eijnden2018}, the optical emission
\citep{Hernandez2019}, and the pulsar properties and accretion torques
\citep{Strohmayer2018a, Bult2021b}. One aspect of this source that has so far
not yet been covered, is its aperiodic variability. The current paper aims to
rectify this situation. The following presents a timing analysis of Neutron
Star Interior Composition Explorer (NICER, \citealt{Gendreau2017}) observations
that were collected between 2017 and 2020.

\section{Data Processing}
\label{sec:data}
The first \nicer observations of \src were collected in 2017 August. Since
then, \nicer has returned to this target at various times, accumulating a total
of 372 ks in unfiltered exposure by 2020 August. These data are available under
the NICER program IDs 103410, 260101, 203410, 303410 and 361201, and were
presented previously in the pulsar timing study of \citet{Bult2021b}. Following
the data cleaning procedures described by these authors, the data were filtered
using \textsc{nicerdas} v7a with standard screening criteria, except for
relaxed constraints on the rate of reset triggers ($<400$ ct\per{s}\per{det})
and on the rate of high energy events ($<1.5$ ct\per{s}\per{det} and $<2.0
\times \textsc{cor\_sax}^{-0.633}$)\footnote{The \textsc{cor\_sax} parameter
models the geomagnetic cut-off rigidity as function of the orbital position of 
the instrument, and is expressed in units of GeV \per{c}.}. This process yielded 202 ks good time
exposure. Barycentric correction were then applied to the clean data based on
the Swift/UVOT position of \citet{AtelRicci08} and the JPL-DE430 ephemeris
\citep{Folkner2014}. 

\section{Analysis and Results}
The \nicer data on \src can be naturally grouped into distinct epochs based on 
their observation dates. For the purpose of this analysis, eight such epochs are
identified. Table \ref{tab:epochs} lists the specifics of these epochs, giving
the grouped ObsIDs with their summed exposures, mean $0.3-9\kev$ count rates, and 
$(0.3-1.5\kev) / (3-9\kev)$ hardness ratio. 

Note that the data grouping used here is slightly different from the one used
in \citet{Bult2021b}. In particular, where they treat the observations in 2018
November and October separately, these data are grouped together into a single
epoch here (epoch 2). Additionally, the data of epoch 7 was collected in
response to an intermediate duration type I X-ray burst and shows large swings in the
source intensity and energy spectrum \citep{Bult2021c}. These swings are seen
in the first $4$ days of monitoring, and are likely due to interactions between
the X-ray burst emission and the accretion flow, which cause disk accretion to
be temporarily inhibited in the aftermath of the burst. Because these
observations do not sample \src in its regular long-term state, the first five
ObsIDs (totalling 15.6 ks) of this set were excluded from the present analysis. 

\begin{table*}[t]
    \movetableright=-0.5in
    \centering
    \footnotesize
    \caption{%
      Data grouping
	\label{tab:epochs}
    }
    \begin{tabular}{l l l c c c}
      \hline \hline
      Epoch  & ObsIDs & Date & Exposure & Rate & Hardness \\
      ~ & ~ & ~ & (ks) & (ct/s) \\
      \tableline
      1 & 1034100101 - 07 & 2017/Aug     &    17.8 & 34.6 & 0.068 \\
      2 & 1034100108 - 22 & 2017/Oct-Nov &    28.5 & 38.3 & 0.081 \\
      3 & 1034100123 - 27 & 2018/Jan     & \phn2.1 & 46.1 & 0.062 \\
      4 & 2601010101 - 04 & 2019/Apr     &    27.5 & 44.3 & 0.066 \\
      5 & 2601010201 - 04 & 2019/Jul     &    23.6 & 35.2 & 0.061 \\
      6 & 2034100101 - 03 & 2019/Oct     &    16.9 & 46.8 & 0.070 \\
      7 & 3034100106 - 12 & 2020/Jun     &    10.3 & 37.4 & 0.055 \\
      8 & 3612010101 - 11 & 2020/Aug     &    60.1 & 37.5 & 0.030 \\
      \tableline
    \end{tabular}
    \flushleft
    \tablecomments{Count rates are calculated in the $0.3-9$ keV band,
    while the hardness ratio is defined as $3-9\kev$ over the $0.3-1.5\kev$
    band rates.
    }
\end{table*}

\subsection{Timing Analysis}
\label{sec:pds}
For the aperiodic timing analysis I investigate the power density 
spectra for each of the eight data epochs. First, the $0.3-9\kev$ event
data were binned into a light curve at $1/8192$-s time resolution and
divided into 64-s duration segments. Each segment was then transformed
into the Fourier domain and converted to a Leahy-normalized power density spectrum
\citep{Leahy1983b} with a frequency resolution of $\approx0.015$ Hz and
a Nyquist frequency of $4096$ Hz. The resulting power density spectra
were then averaged per epoch. The Poisson noise level was estimated by
averaging all powers with frequencies above $2000\hz$. In all cases
the estimated noise level was consistent with the theoretically expected
mean noise power of two. After subtracting the Poisson noise, the power
spectra were finally renormalized to units of squared fractional rms with
respect to the source count-rate. 

The power spectrum of each epoch shows the same overall morphology: below
$10\hz$ the variability is characterized by strong band-limited noise, while
above $10\hz$ the power approaches the Poisson noise level and sometimes showed
weak QPOs. This spectral shape is consistent with the power spectra observed
from hard state neutron star LMXBs \citep{Klis2006}, and identifies the source
state as the (extreme) island state in ``atoll'' naming convention
\citep{Hasinger1989}. To quantify the power spectra, I model them using the
multi-Lorentzian model of \citet{Belloni2002}, with each Lorentzian expressed
in terms of its integrated power, characteristic frequency ($\nu_{\rm max}
= \sqrt{ \nu_0^2 + (W/2)^2 }$, where $W$ and $\nu_0$ are the full width at half
maximum and centroid frequency), and quality factor ($Q = \nu_0 / W$). An individual Lorentzian component is deemed
significant if the ratio of its integrated power (from zero to infinity) over
the lower bound $1\sigma$ uncertainty is greater than three. The best-fit
parameters obtained through this modeling are listed in Table \ref{tab:pds}.

At low frequencies ($<10\hz$), the band-limited noise consistently required two
Lorentzian profiles to obtain a good fit. Using the neutron star naming
convention \citep{Straaten2003, Straaten2005, Altamirano2008b}, the lower
frequency Lorentzian can be identified as the ``break'' component, and higher
frequency term as the ``hump'' component. Epoch 8 showed an additional sharp
QPO at $2.6\hz$. Following the same naming convention, this component is
identified as the low-frequency (LF) QPO. 

At higher frequencies ($>10\hz$), the power spectral shape was less consistent
across the different epochs. The four epochs with the least amount of exposure
did not show any significant power at higher frequencies. The remaining epochs
did show higher frequency power, mostly in the form of an additional broad
noise component. Such broad high frequency noise is not uncommon in the power
spectra of neutron stars, and is usually attributed to either the so-called
$\ell$ow or hectohertz components. The specific identification relies on
observing how the characteristic frequencies of these Lorentzians change in
relation to each other as a function of source flux. Given that such evolution
does not occur in \src (at least not within the considered dataset), a precise
identification is not possible. Instead I simply label these terms as
``high-frequency noise'' (HFN). Epoch 2 stands out for showing a possible QPO
at $330\hz$. With a significance of $2.76\sigma$, however, the feature is not
formally significant. Because QPO amplitudes are expected to increase with
photon energy, I reconstructed the power spectrum in the $1.5-9\kev$ band. Under
this tighter energy selection, the QPO significance increases to $3\sigma$,
hence I tentatively consider it a marginal detection. 

\begin{figure}[t]
  \centering
  \includegraphics[width=\linewidth]{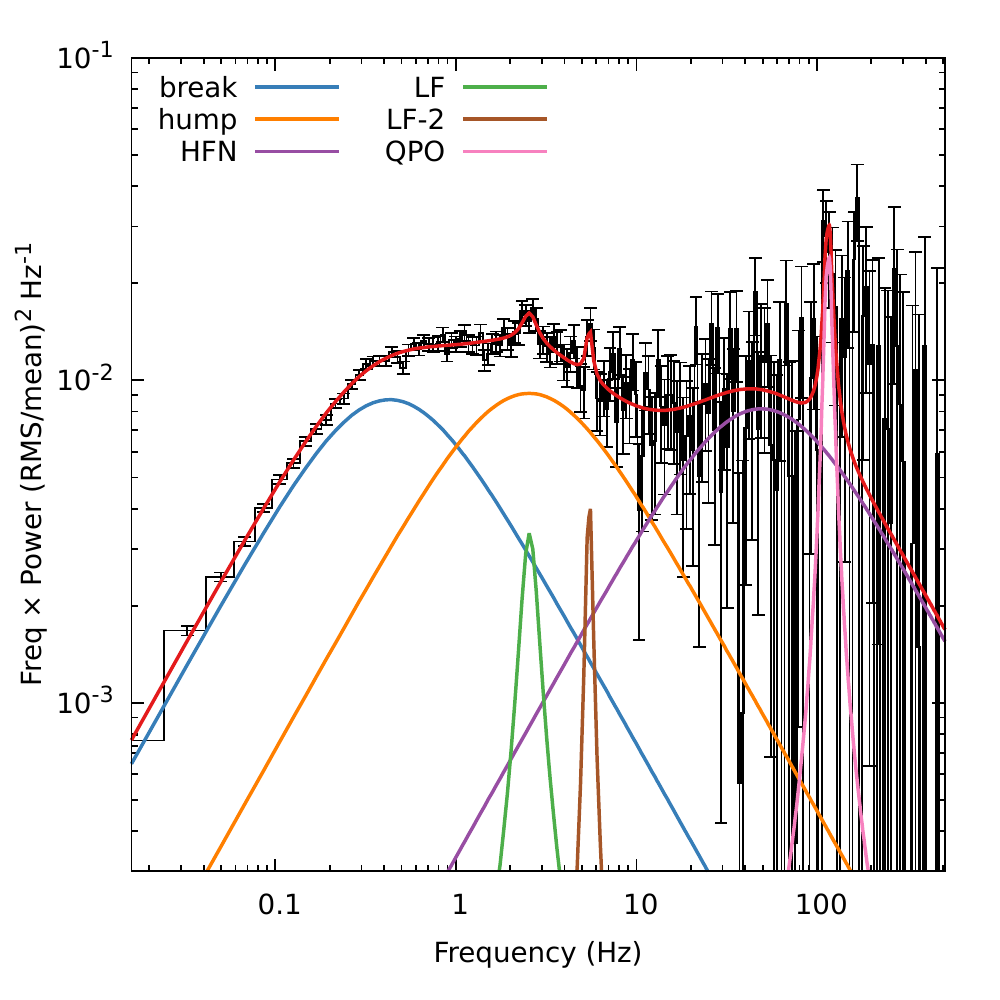}
  \caption{%
    Averaged power density spectrum of \src obtained by combining all data
    except epoch 6 (Table \ref{tab:epochs}). The parameters of the best-fit
    model are listed in Table \ref{tab:pds}. All error bars show $1\sigma$
    uncertainties.
  } 
  \label{fig:pds}
\end{figure}

Comparing the fit parameters of the different epochs, it becomes clear that all
observations except for epoch 6 have a similarly shaped band-limited noise
structure. This similarity suggests that these data can be combined to improve
sensitivity. Indeed, the combined power spectrum is more complex, as shown in
Figure \ref{fig:pds} and reported in the bottom entry of Table \ref{tab:pds}.
In addition to the expected broad noise components and LF QPO, it showed two
previously undetected sharp QPOs: one centered at $5.6\hz$, and the other at $115\hz$. While the
former might be related to the harmonic structure of the LF QPO, the latter is
not easily identified. The possible origins of these QPOs are discussed in Section
\ref{sec:lf qpo}.

\begin{table*}[t]
    \movetableright=-0.5in
    \centering
    \caption{%
      Power spectrum model parameters
	\label{tab:pds}
    }
    \newcommand{\ma}{\tablenotemark{*}}
    \begin{tabular}{l l l l l l}
      \hline \hline
      Epoch &  ID  &  $\nu_{\rm max}$ (Hz) & Quality &  rms (\%) & $\chi^2$/dof \\
      \tableline
      \multirow{2}{*}{1} & break  &  $0.36\pm0.04$  &  0 (fixed)  &  $14.6\pm0.8$ & \multirow{2}{*}{90 / 88}  \\
                         & hump   &  $2.9 \pm 0.5$  &  0 (fixed)  &  $18.8\pm0.7$ & \\
      \tableline
      \multirow{3}{*}{2} & break   &  $0.46\pm0.03$  &  0 (fixed)      &  $17.5\pm0.6$  &  \multirow{3}{*}{66 / 92}  \\ 
                         & hump    &  $3.5\pm0.3$    &  $0.33\pm0.15$  &  $16.2\pm1.1$  &  \\
                         & QPO\ma  &  $330\pm13$     &  $5.3\pm2.7$    &  $14.2\pm2.6$  &  \\
      \tableline
      \multirow{2}{*}{3} & break    &  $0.37\pm0.11$  &  $0.21\pm0.16$  &  $15.7\pm3.1$ &  \multirow{2}{*}{29 / 47}  \\  
                         & hump     &  $2.1\pm1.3$    &  0 (fixed)      &  $16.0\pm2.9$ &  \\
      \tableline
      \multirow{3}{*}{4} & break    &  $0.43\pm0.03$  &  0 (fixed)  &  $17.4\pm0.8$  &  \multirow{3}{*}{92/94} \\
                         & hump     &  $2.4\pm0.3$    &  0 (fixed)  &  $18.0\pm0.7$  &  \\
                         & HFN      &  $40\pm1.1$     &  0 (fixed)  &  $17.6\pm2.1$  &  \\
      \tableline
      \multirow{3}{*}{5} & break    &  $0.50\pm0.04$  &  0 (fixed)  &  $16.7\pm0.7$  &  \multirow{3}{*}{107 / 85}  \\  
                         & hump     &  $3.6\pm0.7$    &  0 (fixed)  &  $17.3\pm0.9$  &  \\  
                         & HFN      &  $104\pm18$     &  0 (fixed)  &  $32.1\pm1.7$  &  \\  
      \tableline
      \multirow{2}{*}{6} & break    &  $0.88\pm0.07$  &  0 (fixed)  &  $18.1\pm0.8$  &  \multirow{2}{*}{98 / 87} \\
                         & hump     &  $6.5\pm1.4$    &  0 (fixed)  &  $16.3\pm0.9$  &  \\ 
      \tableline
      \multirow{2}{*}{7} & break    &  $0.38\pm0.04$  &  0 (fixed)  &  $15.4\pm0.9$  &  \multirow{2}{*}{83 / 88} \\  
                         & hump     &  $3.4\pm0.6$    &  0 (fixed)  &  $19.1\pm0.9$  &  \\
      \tableline
      \multirow{4}{*}{8} & break    &  $0.42\pm0.03$  &  0 (fixed)   &  $16.3\pm0.6$  &  \multirow{4}{*}{65 / 83} \\  
                         & hump     &  $2.5\pm0.3$    &  0 (fixed)   &  $17.1\pm0.8$  &  \\  
                         & LF       &  $2.55\pm0.04$  &  $7.0\pm3.3$ &  $3.9\pm0.8$   &  \\  
                         & HFN      &  $43\pm16$      &  0 (fixed)   &  $14.9\pm1.5$  &  \\  
      \tableline
      \multirow{6}{*}{c} & break  &  $0.425\pm0.015$  &  0 (fixed)    &  $16.5\pm0.4$  &  \multirow{6}{*}{168 / 186} \\
                         & hump   &  $2.5\pm0.2$      &  0 (fixed)    &  $17.0\pm0.5$  &  \\
                         & HFN    &  $48\pm11$        &  0 (fixed)    &  $15.9\pm1.0$  &  \\
                         & LF     &  $2.57\pm0.07$    &  $4.3\pm2.0$  &  $3.5\pm0.8$   &  \\
                         & LF-2   &  $5.56\pm0.12$    &  $13\pm4$     &  $2.6\pm0.4$   &  \\
                         & QPO    &  $115.4\pm2.4$    &  $7.3\pm3.5$  &  $7.0\pm1.3$   &  \\
      \tableline
    \end{tabular}
    \flushleft
    \tablenotetext{*}{Marginal detection.}
    \tablecomments{The last listed epoch (c) gives the model parameters for the
    combination of all data, excluding epoch 6. Uncertainties are quoted at
    $68\%$ confidence.}
\end{table*}

\subsection{RMS energy dependence}
\label{sec:rms}
\begin{figure}[t]
  \centering
  \includegraphics[width=\linewidth]{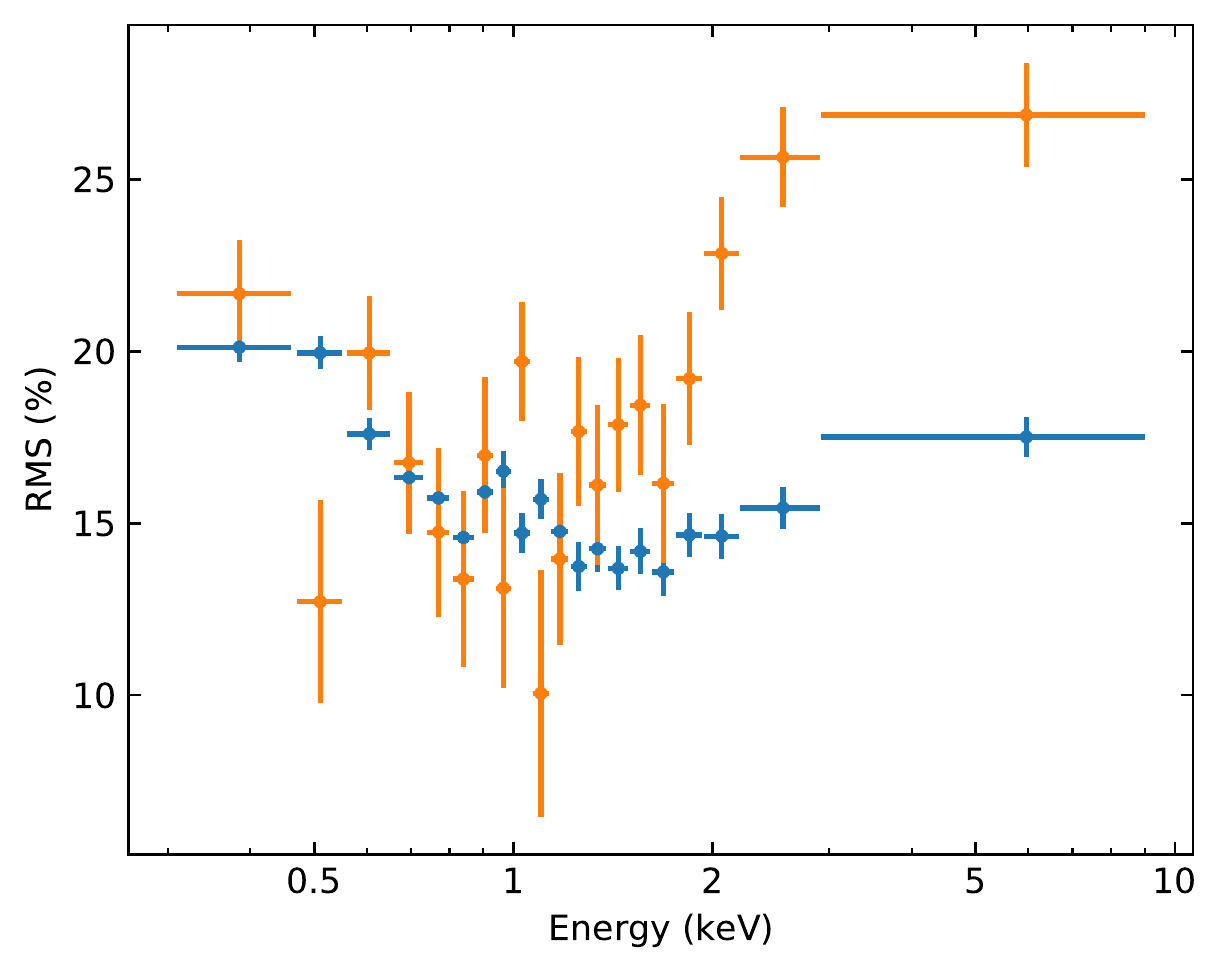}
  \caption{%
    Energy spectrum of the integrated rms in the $1/16-1\hz$ frequency interval
    (blue) and the $1-16\hz$ interval (orange). These intervals were chosen to
    correspond (roughly) with the break and hump components of the
    multi-Lorentzian power spectrum model (see Table \ref{tab:pds}). All
    error bars show $1\sigma$ uncertainties.
  } 
  \label{fig:rms spectrum}
\end{figure}

  To investigate the energy dependence of the band-limited noise, I combined all
  data except for epoch 6 and calculated the rms energy spectrum. The
  $0.3-9\kev$ energy range was divided into $20$ energy bins, such that each bin
  contained a roughly equal number of photons. The average power density
  spectrum was constructed for each energy bin and used to calculate the integrated fractional
  rms amplitudes between $1/16-1\hz$ and $1-16\hz$. These intervals were chosen
  such that they scale geometrically and can be associated (approximately) with
  the break and hump components identified in the energy-averaged power
  spectra. 
  As shown in Figure \ref{fig:rms spectrum}, the rms of the slower variability
  is found to show a minimum at about $1.5\kev$, and increases toward both
  higher and lower photon energies. The higher frequency interval shows a larger
  scatter, but appears to follow the same increasing trend below $1\kev$. Above
  $1\kev$, however, the rms increases more steeply and to a higher amplitude. 

\subsection{Spectral timing}
\label{sec:spectral timing}
A more detailed view of the joint spectral and temporal variations in the
observed data can be obtained through the energy dependent cross spectrum
\citep{Uttley2014}. To compute the cross spectrum, I divide the $0.3-9\kev$
energy range into $70$ subject bands, each containing a roughly equal number of
counts. For each subject band, a 1/128-s resolution light curve was divided the
into 32-s duration segments, which were subsequently transformed into the Fourier
domain. Similarly, an analogously constructed light curve 
in the $0.3-9\kev$ reference band data was also transformed into Fourier space.
The cross spectrum was then constructed by cross-correlating the subject bands
with the reference band and averaging all segments.
The trivial correlation between the subject and reference bands was subtracted
from the cross spectrum and the associated uncertainties were calculated using
the expressions of \citet{Ingram2019b}. Similar to the previous section, the
resulting cross spectrum was integrated in two geometrically spaced frequency
intervals: $1/16-1\hz$ and $1-16\hz$. Finally, the frequency-integrated cross
spectra were normalized to complex covariance \citep{Uttley2014, Ingram2019b}. 

\begin{figure*}[t]
  \centering
  \includegraphics[width=\linewidth]{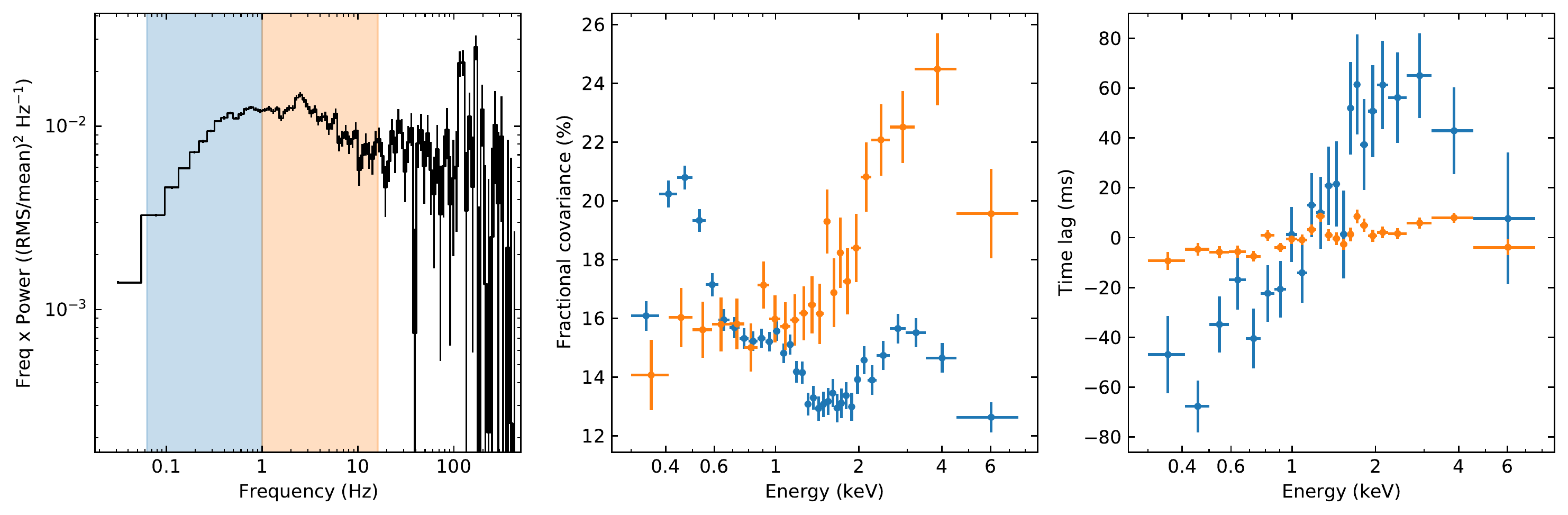}
  \caption{%
    Complex covariance of \srcfull in two frequency intervals. The left panel
    shows the reference band ($0.3-9\kev$) power density spectrum in black,
    with blue and orange bands highlighting the $1/16-1\hz$ and $1-16\hz$
    frequency intervals, respectively. The middle panel shows the fractional
    covariance amplitude for the two frequency intervals as a function of energy.
    The right panel shows the covariant time-lag associated for the two
    frequency intervals as a function of energy. All error bars show $1\sigma$
    uncertainties.
  } 
  \label{fig:cov panels}
\end{figure*}

Figure \ref{fig:cov panels} shows the obtained energy dependent covariance in
the two selected frequency intervals. The left panel highlights frequency
intervals in comparison to the reference band power spectrum, whereas the other
two panels show the fractional amplitude of the complex covariance spectra, and
the associated time-lags (calculated as $\tau = \phi/(2\pi\nu)$, where $\phi$ is the phase
angle of the complex covariance, and $\nu$ the geometric mean of the frequency
band). A few noteworthy properties can be deduced from this figure. 
  First, the covariance pivots as a function of frequency: the faster (hump)
component is less covariant below $1\kev$ and more covariant above $1\kev$ as
compared with the slower (break) component. This indicates that the faster
variations have a harder energy spectrum. 
  Second, the time-lag between the subject and reference bands increases with 
energy. This implies that the stochastic variability is seen in the soft band 
first, and only arrives at harder photon energies at later times. 
  Third, at the highest photon energies both the fractional covariance and the 
lag turn over and start decreasing. This suggests that the complex covariance
is build up from multiple emission components, each intrinsically variable and
slightly out of phase.

\subsubsection{Complex covariance models}
Casting the covariance in units of absolute rms allows the spectrum to be
convolved with the instrument response and to be fit with spectral modeling
tools. Following \citet{Mastroserio2018}, I express the spectral model in the
complex plane, and then jointly fit the real and imaginary projections to the
data using \textsc{xspec} (version 12.11; \citealt{Arnaud1996}). For spectral
models that are time variable in their normalization only, the real line
implementations available within \textsc{xspec} can be embedded into the
complex plane by multiplying them with a factor $e^{i\varphi}$. The real
projection of spectrum $s(E)$, with $E$ the photon energy, thus follows trivially as $s(E) \cos \varphi$,
and the imaginary part as $s(E) \sin \varphi$. If the parameters shaping the
spectral 
density are time-variable as well, however, then the expected shape of the covariance
spectrum will not be the same as the time-averaged version. To illustrate, consider
the well explored example of a power law model that has a time-variable photon
index \citep{Kotov2001, Koerding2004, Mastroserio2018}. Assuming that the
variability is small compared to the time-averaged photon index, one can write
$\Gamma(t)=  \Gamma_0 + \Gamma_1(t)$, where $\Gamma_1(t) \ll \Gamma_0$. To
first order in photon index, the time-dependent energy spectrum can then be
expressed as
\begin{equation}
  s_{\rm pl}(E, t) = a(t) E^{-\Gamma_0} ( 1 + \Gamma_1(t) \log E),
\end{equation}
where $a(t)$ is used to encapsulate the time-variability of the normalization.
Again following \citet{Mastroserio2018}, I can adopt the notation $A(\nu)$ and
$B(\nu)$ for the Fourier transforms of the time variable function $a(t)$ and
$b(t) = a(t)\Gamma_1(t)$, respectively, the Fourier transformed spectral
density becomes
\begin{equation}
  S_{\rm pl}(E, \nu) = A(\nu) E^{-\Gamma_0} + B(\nu) E^{-\Gamma_0} \log E.
\end{equation}
To construct the complex covariance, one has to multiply the above expression
with the complex conjugate of the Fourier transform of the reference band, and
then divide by its modulus. The model covariance then takes the form
\begin{align}
  C_{\rm pl}(E, \nu) 
    = A(\nu) &\Bigg[ e^{i \varphi_A(\nu)} E^{-\Gamma_0} \nonumber \\
    &+ \rho(\nu) e^{i \varphi_B(\nu)} e^{-\Gamma_0} \log E \Bigg], \label{eq:covpl}
\end{align}
where $\varphi_A(\nu)$ and $\varphi_B(\nu)$ give the frequency dependent phase
lags between the subject and reference bands associated with the first and
second order terms of the Taylor expansion in photon index, and $\rho(\nu)
= |B(\nu)| / |A(\nu)|$ gives a correlation factor between these two terms. While
one could attempt relate the $\rho(\nu)$ and $\varphi_B(\nu)$ more directly to 
the variability of the photon index, this would require adopting a physical model
for both $a(t)$ and $\Gamma_1(t)$. As noted by \citet{Mastroserio2018}, such
a model assumption is not actually required for fitting the covariance. Thus,
for an integrated frequency range, the complex covariance of a variable power
law is a five parameter model that depends on the time-averaged photon index,
$\Gamma_0$, and on on four additional Fourier terms: $A$, $\rho$, $\varphi_A$,
and $\varphi_B$.

In a similar vein, I can also derive the complex covariance of a blackbody emitter that
varies in both normalization and temperature. Consider that the spectral density of
blackbody radiation is
\begin{equation}
  s_{\rm bb}(E, t) = a(t) \frac{E^2}{\exp(E/ T(t)) - 1},
\end{equation}
where $T(t)$ gives the time-variable blackbody temperature is units of energy, and
the physical constants have been absorbed into the time-variable normalization 
$a(t)$. Again, assuming that the variability in temperature is small compared
to the time-averaged mean, one can write $T(t) = T_0 + T_1(t)$ with $T_1(t) \ll
T_0$. Then, the first order Taylor expansion of the blackbody spectrum around
$T_0$ is
\begin{align}
  s_{\rm bb}(E, t) 
  &= a(t) \Bigg[ \frac{E^2}{\exp(E/T_0) - 1} \nonumber \\
  &+ T_1(t) \frac{E^3 \exp(E/T_0)}{ T_0^2 (\exp(E/T_0) - 1)^2} \Bigg],
\end{align}
Through an analogous construction as applied in the previous paragraph,
the complex covariance becomes
\begin{align}
  C_{\rm bb}(E, \nu) &= A(\nu) \Bigg[ e^{i \varphi_A(\nu)} \frac{E^2}{\exp(E/T_0) - 1} \nonumber \\
  &+ \rho(\nu) e^{i \varphi_B(\nu)} \frac{E^3 \exp(E/T_0)}{ T_0^2 (\exp(E/T_0) - 1)^2} \Bigg],
  \label{eq:covbb}
\end{align}
where, for ease of notation, I have recycled the naming convention of the four
Fourier terms ($A, \rho, \varphi_A, \varphi_B$). Equations \ref{eq:covpl} and
\ref{eq:covbb} were implemented as local \textsc{xspec} models and have been
made publicly available\footnote{\url{https://github.com/peterbult/cov-models}}.
In the following, I will refer to these models as \texttt{covpl} and
\texttt{covbb}, respectively.

\subsection{Spectral fitting}
The time-averaged spectrum of \src is well-described as an
absorbed blackbody and power law, with an additional Gaussian emission line
centered at $1\kev$ \citep{Degenaar2017, Eijnden2018, Bult2021b}. Adopting the
T\"ubingen-Boulder model for the interstellar absorption \citep{Wilms2000}, the
analogous spectrum for the complex covariance can be expressed as
\begin{quote}
  \texttt{tbabs $\times$ (complex$\times$gauss $+$ covbb $+$ covpl)},
\end{quote}
where \texttt{complex} is a simple multiplicative \textsc{xspec} model that
calculates either the sine or cosine of a joint phase argument, depending
on a switch parameter. 

Applying the above model without any external constraints produces
a reasonable good fit to the complex covariance spectrum of \src, yielding
a best-fit $\chi^2/{\rm dof}$ of $182/122$ for the lower frequency interval and
$152/122$ for the higher frequency interval. The associated best-fit parameters
are listed in Table \ref{tab:cov}.

%

\begin{table*}[t]
  \newcommand{\twoline}[2]{\begin{tabular}{@{\hspace{1em}}l@{}}#1 \\ #2\end{tabular}}
    \centering
    \movetableright=-0.5in
    \caption{%
      Covariance spectrum model parameters
	\label{tab:cov}
    }
    \newcommand{\ma}{\tablenotemark{*}}
    \begin{tabular}{l l@{\hspace{0.5in}} l l l@{\hspace{0.5in}} l l l}
      \hline \hline
      ~ & ~         &  \multicolumn{2}{c}{Free}  & & \multicolumn{3}{c}{Tied}  \\ 
      \tableline
      Component & Parameter &  break & hump  & & break & hump & time-averaged\\ 
      ~ & ~ & $(1/16-1\hz)$ & $(1-16\hz)$ & & $(1/16-1\hz)$ & $(1-16\hz)$ \\
      \tableline
      \multirow{5}{*}{covbb}     &  kT (keV)            &   $0.61_{-0.07}^{+0.06}$     &  $0.62_{-0.04}^{+0.12}$  & &  \multicolumn{3}{c}{$0.4049_{-0.0020}^{+0.0026}$}         \\
				 &  $\varphi_A$ (deg)   &   $11_{-8}^{+7}$             &  $8_{-10}^{+11}$         & &  $-30\pm20$                 &  $-60\pm30$              &  -   \\
				 &  $\varphi_B$ (deg)   &   $179_{-16}^{+12}$          &  $160\pm20$              & &  $39_{-12}^{+4}$            &  $33_{-8}^{+14}$         &  -   \\
				 &  $\rho$	        &   $0.10_{-0.05}^{+0.04}$     &  $0.09_{-0.03}^{+0.08}$  & &  $0.16_{-0.04}^{+0.09}$     &  $0.24\pm0.10$           &  -   \\
				 &  K$_{\rm bb}$ (km/10 kpc)$^2$ &   $2.5\pm0.7$                &  $4.3_{-1.7}^{+0.5}$     & &  $3.4_{-1.1}^{+0.2}$        &  $4.8\pm1.4$             &  $92_{-4}^{+2}$   \\
      \tableline 
      \multirow{5}{*}{covpl}     &  $\Gamma$            &   $1.95_{-0.07}^{+0.06}$     &  $1.73_{-0.14}^{+0.30}$  & &  \multicolumn{3}{c}{$1.786_{-0.007}^{+0.019}$}            \\
				 &  $\varphi_A$ (deg)   &   $-3.8_{-1.3}^{+1.8}$       &  $-10_{-3}^{+6}$         & &  $-2.9_{-0.8}^{+1.0}$       &  $-3.9_{-3.0}^{+1.6}$    &  -   \\
				 &  $\varphi_B$ (deg)   &   $-100_{-50}^{+140}$        &  $-160\pm30$             & &  $-22_{-10}^{+7}$           &  $-159_{-9}^{+14}$       &  -   \\
				 &  $\rho$              &   $0.03_{-0.03}^{+0.05}$     &  $0.11_{-0.10}^{+0.28}$  & &  $0.126_{-0.018}^{+0.020}$  &  $0.22_{-0.06}^{+0.09}$  &  -   \\
				 &  norm ($\E{-3}$)     &   $2.19_{-0.11}^{+0.09}$     &  $2.10_{-0.13}^{+0.31}$  & &  $1.91_{-0.03}^{+0.09}$     &  $1.90_{-0.11}^{+0.16}$  &  $8.09_{-0.09}^{+0.24}$ \\
      \tableline 
                                &  line energy (keV)   &   $1.002_{-0.016}^{+0.013}$  &  $0.96_{-0.04}^{+0.06}$  & &  \multicolumn{3}{c}{$0.970_{-0.004}^{+0.005}$}             \\
      complex$\times$           &  width (keV)         &   $0.136_{-0.012}^{+0.019}$  &  $0.15_{-0.05}^{+0.02}$  & &  \multicolumn{3}{c}{$0.172_{-0.008}^{+0.005}$}             \\
      Gaussian                  &  $\varphi_A$ (deg)   &   $-2_{-7}^{+8}$             &  $10_{-14}^{+10}$        & &  $1_{-6}^{+2}$               &  $13_{-13}^{+12}$        &  -   \\
				&  norm ($\E{-4}$)     &   $3.0_{-0.3}^{+0.5}$        &  $3.2_{-1.1}^{+0.5}$     & &  $3.9\pm0.3$                 &  $3.7_{-0.8}^{+1.0}$     &  $18.1_{-0.9}^{+0.8}$   \\
      \tableline
      &  $\chi^2$ / dof      &   $182 / 122$ & $152 / 122$ & & \multicolumn{3}{c}{$584 / 464$}  \\
      \tableline
    \end{tabular}
    \flushleft
    \tablecomments{All phases express the lag with respect to the $0.3-9\kev$ reference band. 
    The absorption column density was set to $N_H = 0.1\E{22}\persq{cm}$ \citep{Bult2021b}.
    Uncertainties are quoted at $90\%$ confidence.}
\end{table*}

Compared to the spectral parameters of the time-averaged spectrum, 
the complex covariance yields a larger mean blackbody temperature (0.6\kev
versus 0.4\kev). 
In an alternative approach, I also fit the covariance spectra jointly with
the time-averaged energy spectrum, which allows for the mean blackbody temperature,
the power law photon index, and the Gaussian line parameters of the covariance spectra to be tied to their respective
counterparts in the time-averaged spectrum.  Even with the mean spectral shape
constrained, the covariance model is still able to adapt to the data. The total
$\chi^2$ of this fit is $584$ for $464$ degrees of freedom. The specific $\chi^2$
contribution of the lower frequency interval covariance is $188$, while the
higher frequency interval contributed a $\chi^2$ of $155$. Hence, this
alternative description of the covariance data is slightly worse, but not
significantly so. Figure \ref{fig:cov model} shows the best-fit model and
residuals in the real and imaginary projection, and the detailed best-fit
parameters are listed in Table \ref{tab:cov}.

\begin{figure*}[t]
  \centering
  \includegraphics[width=\linewidth]{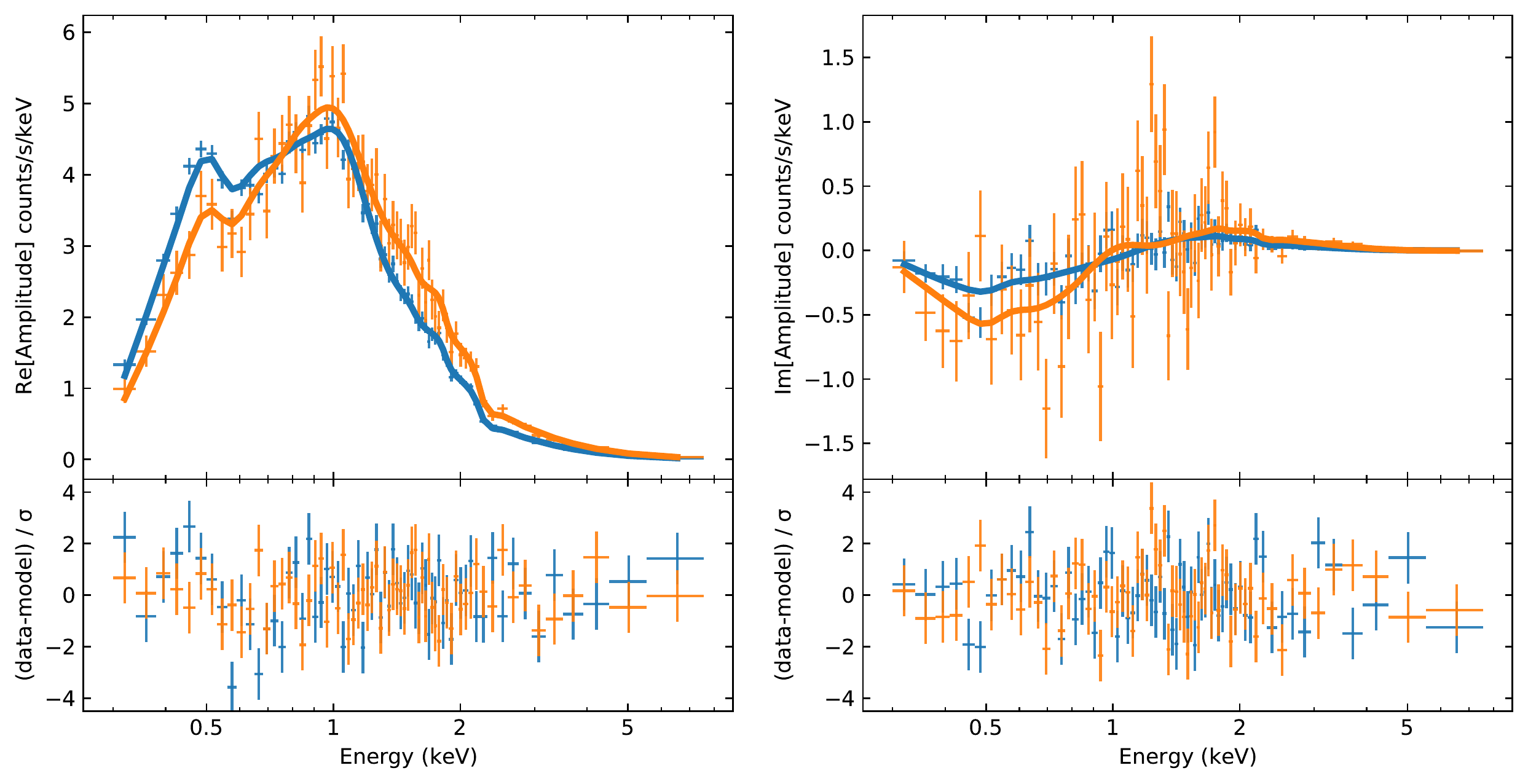}
  \caption{%
    Model fit to the complex covariance in the real (left) and imaginary (right)
    planes (see Table \ref{tab:cov}, `tied' model). The top panels show the data and
    best fit models, while the bottom panels show the residuals divided by the uncertainties.
    In all panels the blue curve and data show the $1/16-1\hz$ frequency
    interval, while the orange curve and data show the $1-16\hz$ interval. All
    error bars show $1\sigma$ uncertainties.
  } 
  \label{fig:cov model}
\end{figure*}

\section{Discussion}
\label{sec:discussion}
I have presented an analysis of the stochastic variability of \src using \nicer
observations. A study of the power density spectrum showed that the aperiodic
variability is (mostly) stable with time, showing pronounced band-limited noise
as well as several QPOs. A spectral analysis of the energy-dependent complex
covariance further showed that the band-limited noise can be described with the
same blackbody plus power law model explaining the time-averaged energy
spectrum when suitably accounting for the time-dependence of the blackbody
temperature and power law photon index. In the following I first briefly
consider the one epoch that showed a deviating power spectrum, and follow with
a discussion of the main results relating to the observed QPOs and the
energy-dependent complex covariance analysis of the band-limited noise. 

\subsection{Epoch 6}
All epochs analysed in this paper showed power spectra dominated by broad
band-limited noise with $\sim25\%$ rms below $16\hz$. Such power spectra are
typical for (hard) island state accreting neutron stars \citep{Klis2006}, and
could be well described with two Lorentzian profiles: the break and hump
components. For most epochs, the shape of this broad noise was nearly identical,
as evident from the tight clustering of the power spectral fit parameters
listed in Table \ref{tab:pds}. Only epoch
6 notably deviates, with break and hump frequencies that are each about
  a factor two larger than in the other epochs. 

It is not at all clear why only epoch 6 deviates from the larger sample.
Systematic shifts in the characteristic frequencies are common in low mass
X-ray binaries \citep{Wijnands1999, Psaltis1999a}, but should correlate with
spectral hardness or luminosity as the evolution is believed to be driven by
a changing mass accretion rate. Considering the position of this one epoch in
terms of its count-rate and hardness ratio (Table \ref{tab:epochs}), however,
it is clearly not an outlier. Of course the correlated temporal and spectral
source state evolution is observed over orders of magnitude changes in
characteristic frequencies and source intensity. The observed change in
frequencies during epoch 6 are only a factor two, so perhaps the underlying
change in the variability mechanism (presumably an increase in mass flow
through the disk) was simply not strong enough or sustained for long enough to
trigger a similar change in the emission processes.

\subsection{Quasi-periodic oscillations}
\label{sec:lf qpo}

The power spectra of \src show two QPOs at low frequencies. The first is
centered at $2.6\hz$ and was identified as the LF QPO in the variability
nomenclature for neutron stars. The QPO is observed in epoch 8 and in the
combined power spectrum, although in the latter case the feature is slightly
broader. Such broadening is not uncommon when averaging power spectra over many
different observational epochs \citep[see, e.g.,][for a discussion]{Doesburgh2017}, and suggests that the
underlying QPO frequency has a modest scatter across the different epochs. The
second QPO is centered at $5.6\hz$ and seen in the combined power spectrum
only. The centroid frequency of this second QPO is consistent with being an
integer multiple of the LF QPO centroid frequency (at a $2\sigma$ level),
suggesting that the two QPOs might be harmonically related. Higher harmonics of
the LF QPO are again a common feature in the power spectra of neutron stars,
especially in the hard state \citep{Doesburgh2017}. The one aspect of these QPO
that is unusual is their frequency relation with respect to the hump component.
In a comprehensive study of LF QPOs across various accreting neutron stars
\citet{Doesburgh2017, Doesburgh2019} found that the LF QPO frequency is always
smaller than that of the hump, while the frequency of the LF second harmonic is
about the same as the hump frequency. In contrast, the LF frequencies seen in
\src are about once and twice the hump frequency, suggesting that either \src
does not neatly follow the established frequency ratios, or that the observed
QPOs are really the second and fourth harmonics.

It has long been recognized that the frequency relation between the neutron
star LF QPO and the band-limited noise has a striking commonality with the
frequency relations between type C QPOs and band-limited noise observed from
black hole binaries \citep{Psaltis1999a, Wijnands1999}. Hence, it is commonly
assumed that the two features share a common origin that is independent of the
precise nature of the compact object. Studies of the typically much more
pronounced type C QPOs has provided strong evidence that driving mechanism is
geometric in nature \citep{Motta2015, Heil2015, Eijnden2017}. While various
models have been proposed to explain these QPOs \citep[see][for a detailed
review]{IngramMotta2019}, the most sophisticated model associates the QPO with
the solid-body precession of the inner accretion disk \citep{Ingram2009,
Ingram2016}. Similar precession should also occur in accreting neutron star
binaries, although the extrapolation is not straightforward
\citep{Altamirano2012, Bult2015b, Doesburgh2017}, with the stellar oblateness
and dynamically relevant magnetosphere introducing additional torques on the
disk which may modify or even dominate the net precession frequency
\citep{Lai1999, Morsink1999, Shirakawa2002}.  

Another relevant result of type C QPOs is that its second harmonic tends to
have a softer spectrum than both the fundamental and time-averaged emission
\citep{Axelsson2014, Axelsson2016}. If this property generalizes to neutron
stars, it might explain why \nicer, with its comparatively soft passband, would
be more sensitive to the higher harmonics of the LF QPO than its fundamental.
The limited signal power in the QPO unfortunately prevents a more detailed
investigation at this time. 

\subsection{The high-frequency QPOs}
\label{sec:hf qpo}
Two QPOs are observed at higher frequencies: a $115\hz$ QPO detected in the
combined power density spectrum, and a (marginal) $330\hz$ QPO seen in epoch
$2$ only. The fact that these QPOs are the fastest aperiodic features in the
power spectrum and have comparatively high quality factors is reminiscent of
kilohertz QPOs \citep{Klis2006}. Such kHz QPOs can appear as twin peaks,
although regularly only the upper or lower term of the pair is visible.
A ubiquitous property of these QPOs, however, is that their coherence (i.e. the
width of the QPO) increases with luminosity. The sharply pronounced (twin) kHz
QPOs are seen at high luminosity, when the source is on the banana branch (i.e.
the intermediate to soft state). As luminosities decrease, the QPOs broaden and
shift to lower frequencies. In this context, the two high frequency QPOs seen
in \src do not seem consistent with kHz QPOs. Based on the luminosity, energy
spectrum, and the overall shape of the power spectrum, \src is firmly in the
(hard) island state. Indeed, comparing the frequencies of the break and hump
noise components with the frequency relations of other LMXBs \citep[see,
e.g.,][]{Doesburgh2017} implies that \src is located at the
hardest end of the evolutionary track. In this state, upper kHz QPO is expected
to have devolved into incoherent noise, while the lower kHz QPO is either not
visible at all, or perhaps responsible for the so-called $\ell$ow noise
component \citep{Psaltis1999a}. So, while the frequencies of the two observed QPOs are at least
nominally consistent with kHz QPOs, their observed quality factors are normally
only seen at much higher luminosities.

The QPOs do seem to have special frequencies with respect to the known $164\hz$
neutron star spin frequency. Specifically, the $115\hz$ QPO is within
$2\sigma$ deviation of a 2:3 frequency ratio relative to spin frequency. The $330\hz$, meanwhile, is centered at
twice the spin frequency. Note, though, that the latter QPO is not an artifact
of the pulse harmonic: with an rms amplitude of about $14\%$ the QPO is far
stronger than the $1.8\%$ rms amplitude of the pulse waveform
\citep{Bult2021b}, let alone its second harmonic in isolation. Possibly, then,
the rotating magnetosphere is exciting resonance frequencies in disk. In the
case of the $115\hz$ QPO, this resonance might have become visible only due to
the unusual long-term stability of \src that allowed for large amounts of data
to be averaged. The $330\hz$ QPO, seen only in epoch 2, is more difficult to
understand, as the higher frequency must exists somewhere in the disk to be
excited by a resonance mechanism. The fastest periodicity within the accretion
disk should be the orbital frequency at the inner edge of the disk, hence, the
simple detection of a $330\hz$ QPO would imply that the orbital motion must be
at least that fast. Assuming Kepler orbital motion around a canonical $1.4$ solar
mass neutron star with $10\km$ radius, this QPO frequency implies an inner
disk radius of about $36\km$, notably smaller than the $45\sim50\km$ derived
from an analysis of the long-term accretion torque \citep{Bult2021b}. This
discrepancy might imply that the accretion torque modelling was incomplete, or, 
perhaps more plausibly, indicate that the marginal QPO detection was spurious. 

\subsection{Spectral timing}
The energy-dependent complex covariance was calculated in two integrated
frequency intervals that roughly coincide with the break and hump Lorentzian
components of the power spectra. These covariance spectra showed systematic
hard lags, and fractional amplitudes that become harder as the variability
frequency increases. These patterns are qualitatively similar to the hard state
covariance spectra of the neutron star system Aql X-1 \citep{Bult2018a} and the
black hole binary GX 339--4 \citep{Uttley2011}. The key points noted by these
studies are that the softest photon energies (around $0.5\kev$) have high
fractional variability and lead the correlated variations seen at higher
energies. The common interpretation of these results attributes the physical
process driving the band-limited noise to mass accretion rate fluctuations that
propagate inward through the accretion disk \citep{Lyubarskii1997}. The
variability is then observed from the accretion disk first. After some
propagation delay, the variability is transferred to Comptonizing medium, so
that it is observed at higher photon energies at later times. Interestingly, this
model does not readily apply to \src. Multi-wavelength spectral energy density
modelling of \src indicates that the accretion disk is visible mainly at longer
wavelengths and does not meaningfully contribute to the X-ray spectrum
\citep{Hernandez2019}, hence the leading variability at the softest energies
cannot be attributed to thermal disk emission.
In an alternate interpretation of the complex covariance spectra of Aql X-1,
\citet{Bult2018a} suggest that a scenario akin to the QPO model of
\citet{Miller1992, Lee1998} can produce the same spectral-timing characteristics. 
That is, the electron scattering optical depth of the Comptonizing medium may be 
modulated stochastically, for instance by propagating mass accretion rate 
fluctuations. This causes the emergent power law photon index to vary, which can
give rise to energy dependent time lags.

Various studies of \src have found that the time-averaged energy spectrum can
be well described with a simple phenomenological model consisting of an
absorbed blackbody component, a power law, and Gaussian emission line centered
at $1\kev$ \citep{Degenaar2017, Eijnden2018, Bult2021b}. The blackbody
component is generally interpreted as emission from the neutron star stellar
surface, while the power law indicates the presence of a Compton scattering
medium. High resolution spectroscopy has shown that the $1\kev$ emission
consists of a continuum of narrow line features \citep{Degenaar2017,
Eijnden2018}, suggesting an origin in a remote disk reflection site, or
possibly an outflow. In Section \ref{sec:spectral timing} I generalized this
model to be applicable to the complex covariance spectra, and then applied the
model to the observed data. Initially, the model was applied to the covariance
data only, allowing all parameter to be freely optimized. A notable result of
this fit, however, was a mean blackbody temperature of $0.6\kev$, which is
hotter than the $0.4\kev$ seen in the time-averaged emission. This discrepancy
suggests that the thermal emission might be a multi-temperature blackbody. The
two blackbodies could, for example, be associated a cooler stellar surface and
hotter boundary layer. The boundary layer, being fed by the variable mass
accretion flow, could naturally produce the observed (thermal) variability. As
matter spreads out over the stellar surface it is likely the variability
dampens out, such that the brighter surface emission does not show up in the
covariance spectrum. A problem with this interpretation, however, is that it is
inconsistent with the time-averaged energy spectrum. Adding a secondary
blackbody to the spectrum with its temperature fixed to $0.6\kev$, the
associated normalization is found to be $4\pm1 (\km/10 \kpc)^2$. Hence, this
secondary blackbody would have to be modulated at $\approx100\%$ rms to explain
the covariance, which seems improbable. 

\begin{figure*}[tb]
  \centering
  \includegraphics[width=\linewidth]{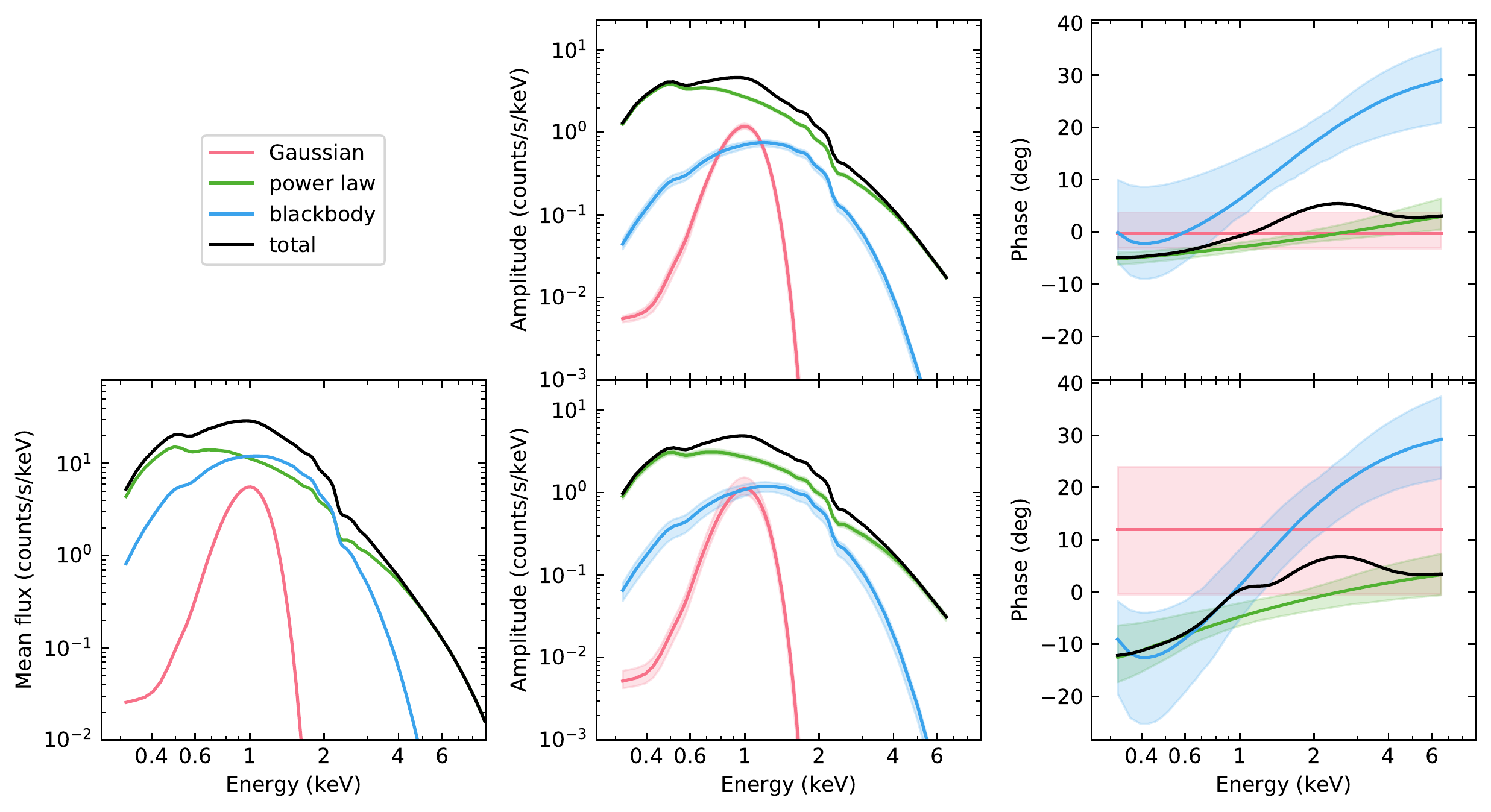}
  \caption{%
    Decomposition of the (folded) complex covariance model (see Table \ref{tab:cov}, 
    `tied' model). The left panel
    shows the time-averaged energy spectrum of \src. The middle column shows the
    amplitude of the complex covariance model, while the right column shows the
    phase. The top and bottom rows show the $1/16-1\hz$ and $1-16\hz$ intervals,
    respectively.
    In all panels the black line shows the best-fit total model, whereas the red, green,
    and light blue lines show the respective contributions from the $1\kev$ Gaussian line, the power
    law, and the blackbody. The shaded areas show the 90\% confidence region associated with each
    spectral component.
  } 
  \label{fig:cov model}
\end{figure*}

In an alternative approach, the complex covariance was also fit jointly with the
time-averaged spectrum so that the time-averaged spectral parameters could be
tied between them. This approach gave a statistically equivalent description of
the covariance spectrum, but has the advantage that it enforces self-consistency.
Considering the best-fit parameters obtained with this approach (as listed in
Table \ref{tab:cov}) one sees that: first, the blackbody varies comparatively
little, with an amplitude of about $5\%$ rms in its normalization; second, the
power law varies by about $23\%$ rms, and appears to be the primary source of
the variable emission; third, the Gaussian emission feature varies at $20\%$,
comparable to the power law. 
The spectral shape of both the blackbody and power law vary as well, but
because no specific model for the time-dependence has been assumed, these
variations cannot be straightforwardly interpreted in terms of $\Gamma(t)$
or $T(t)$. Instead, the fit parameters indicate how much of the fractional rms in the
covariance spectrum is contributed by the shape variations, which is $1\%$ for
the blackbody temperature, and $5\%$ for the power law photon index.
Due to the variable shape of the covariance components, the phase parameters
are similarly difficult to interpret directly. A more approachable way to
inspect the phases is to decompose the model into its additive components
and visualize them in polar coordinates, as shown in Figure \ref{fig:cov model}.
This decomposition directly shows how the phase lags of the individual components
depend on photon energy, and demonstrate that the variability of both the
blackbody and the Gaussian emission feature lag behind the power law. 

Before trying to interpret the spectral fit results in physical terms, it
is important to remember the limitations of the phenomenological spectral
model. Unlike the simple power law model, the emission spectrum of a Compton
scattering medium should roll off at low photon energies \citep{Shapiro1976,
Sunyaev1980}. Considering the time-averaged spectral model shown in Figure
\ref{fig:cov model} (left panel), is should be clear that the power law is the
dominant spectral component below about $1\kev$. Hence, relative to a Compton
scattering spectrum there is an excess of soft flux. A natural explanation for
this soft excess is that it is caused by a disk reflection (see \citet{Keek2017}
and \citet{Bult2021c} for disk reflection modelling of \src). In this sense, the
phenomenological model still makes sense if the bulk of the reflection emission
originates from close to the neutron star. The observed time lags intrinsic to
the power law emission are then much larger than the crossing time delays of
the reflection, so that the low energy region, where the reflection dominates
the total emission, should closely track the variability of the power law. 

Presuming that the phenomenological fit (as shown in Figure \ref{fig:cov model})
is indeed a reasonable approximation to the physical process, the fit results
seem well matched by an oscillating Compton scattering medium \citep{Miller1992,
Lee1998}, as proposed for Aql X-1. The power law component component clearly
dominates the observed variability in \src across photon energies, and its pivoting
photon index induces the observed hard lags. The lag between the Gaussian line
and the power law is easy to understand, as the line feature has been proposed
to originate in an outflow or from a distant disk reflection site. In either case 
the line will be excited by the illuminating power law emission, but have a longer light
travel path. So, light travel time delays are naturally expected for this line.
The lag between the thermal blackbody emission and the power law can also be
explained through one of two scenarios. First, if the electron scattering
optical depth is modulated by propagating mass accretion rate fluctuations,
then these fluctuations might simply be affecting the Compton scattering medium
before they propagate down to the stellar surface. Hence, the time delay
between the power law and blackbody variability could simply be a propagation
delay. Second, as proposed in more recent theoretical implementations of the
variable Compton scattering model \citep{Kumar2014, Karpouzas2020}, the power law
emission of Compton scattering medium should also illuminate the disk and star
giving rise to a radiative feedback mechanism between the star/disk and the
Compton scattering region. Hence, the delayed (and reduced) variability
observed in the blackbody might be induced by the X-ray heating of the stellar surface
by the non-thermal emission. 
Qualitatively, this scenario appears to fit well the spectral modelling of the
covariance spectrum in \src. The next question, then, would be if the model
of \citet{Kumar2014} and \citet{Karpouzas2020}, which is developed for QPOs,
is able to account for the band-limited noise as well, and if the
predictions of that model can quantitatively match the measured covariance spectra of \src.
Given that the aim of this paper was to report on stochastic variability of
\src, this more detailed modelling effort is left for future work. 

\section{Summary}
In summary, I have presented an analysis of the stochastic X-ray variability
of the accreting millisecond X-ray pulsar \src. Over the course of three years
of monitoring, the power density spectra of this source have remained mostly
stable, showing broad band-limited noise at low frequencies, as is typical for
hard state neutron star LMXBs. By combining these data I have found evidence
for a relative common low-frequency QPO as well as an unusual $115\hz$ QPO
that may be due to a 2:3 resonance with the stellar spin frequency. Further,
a spectroscopic study of the complex covariance showed that the continuum
noise can be qualitatively explained as being due to a time variable
Compton-scattering medium, in which the variability observed in the emission
feature and at the lowest photon energies are attributed to the disk reflection
of the power law continuum.

\facilities{ADS, HEASARC, NICER}
\software{heasoft (v6.27.2), nicerdas (v7a)}

\begin{acknowledgments}
This work was supported by NASA through the Astrophysics Data Analysis Program
(grant number 80NSSC20K0288) and the CRESST II cooperative agreement
(80GSFC21M0002), and made use of data and software provided by the High Energy
Astrophysics Science Archive Research Center (HEASARC). 
\end{acknowledgments}

\bibliographystyle{fancyapj}

\end{document}